# Quasi-continuous-wave-pumped thulium-doped fiber laser with 0.2 kW of instantaneous output power

CHANGSHUN HOU,[1, *] ZIWEI ZHAI,[1] NILOTPAL CHOUDHURY,[1] TOM HARRIS,[1] JAYANTA K. SAHU,[1] AND JOHAN NILSSON[1]

[1]*Optoelectronics Research Centre, University of Southampton, Southampton SO17 1BJ, UK*
*\*C.Hou@soton.ac.uk*

**Abstract:** A thulium-doped fiber laser operating quasi-continuous-wave generated 198 W of instantaneous output power in 0.2-ms pulses at 50 Hz repetition rate. The duty cycle becomes 1% and the average output power 2.0 W. This was cladding-pumped with 408 W from 0.79-µm diode lasers (average pump power 4.1 W). The pump switch-on time was ~10 µs at full power, with the laser exhibiting relaxation oscillations starting ~5 µs after the start of the pumping and lasting for ~5 µs. The slope efficiency with respect to absorbed pump power was 67% up to 200 W of absorbed pump power and 52% at the full 352 W of absorbed pump power. The low duty cycle simplified the heatsinking, despite the high instantaneous power and thermal load.

## 1. Introduction

Cladding-pumped thulium-doped fiber lasers are attractive for their power scalability and the availability of high-power, high-brightness 0.79-µm diode-lasers for pumping. Their output power has reached over 1 kW at wavelengths around 2 µm [1]. A high concentration is generally used in order to improve the efficiency through a "two-for-one" cross-relaxation process [2]. This however leads to a high pump absorption and short device lengths. The thermal load per unit length becomes high and leads to coating failure in the absence of highly capable thermal management solutions. Measures such as off-peak pumping can reduce the pump absorption and thus distribute the heat over a longer fiber, even if this reduces the efficiency due to a relatively high background loss at 2 µm [3]. An alternative approach is to operate quasi-continuous-wave (QCW).

In this paper, we report a quasi-continuous-wave thulium-doped fiber laser (TDFL) with 0.2 kW of instantaneous output power. An all-fiber laser cavity was formed by perpendicular cleaves at both ends of the fiber system, leading to double-ended output with approximately half of the power exiting each end of the fiber. The thulium-doped fiber (TDF) was cladding-pumped with up to 0.4 kW of instantaneous power at 0.79 µm from two diode lasers and launched through a pump/signal combiner spliced to the TDF. A QCW laser diode driver (LDD) generated 0.2-ms pulses to drive the diode lasers in series with up to 16.9 A of current. The heatsinking of the TDFL (including the diode lasers) was water-free and the TDF was simply lying in a loose coil on the optical table.

## 2. Experimental setup

Fig. 1 illustrates the experimental setup of the TDFL. The TDF was designed and fabricated in-house with modified chemical vapor deposition (MCVD) and solution-doping. The pump is confined within the quasi-octagonally shaped silica inner cladding by a low-index coating. This was a standard acrylate coating, i.e., not a high-temperature version. The TDF was cladding-pumped by two 0.79-µm fiber-coupled diode lasers (BWT) launched via a (2+1) x 1 pump and signal combiner (DK Photonics) spliced to the TDF. The diode lasers were driven in series by a LDD (Meerstetter), capable of fast-pulse operation (on & off switching times < 10 µs). The LDD was controlled by a waveform generator (Tektronix) via a current control port on the LDD to produce rectangular pulses of 200-µs duration at a pulse repetition frequency (PRF) of 50 Hz (duty cycle 1%). The overall temporal characteristics of the TDFL system followed that of the diode driver. Each pump diode is rated for 130 W of continuous-wave (CW) power in a 105-µm core fiber with a NA of 0.22. In QCW operation, it can be driven to produce around 220 W of instantaneous pump power in 200-µs pulses. The measured pump transmission efficiency of the combiner is around 90%. Therefore, 408 W of maximum instantaneous pump power was launched into the gain fiber.

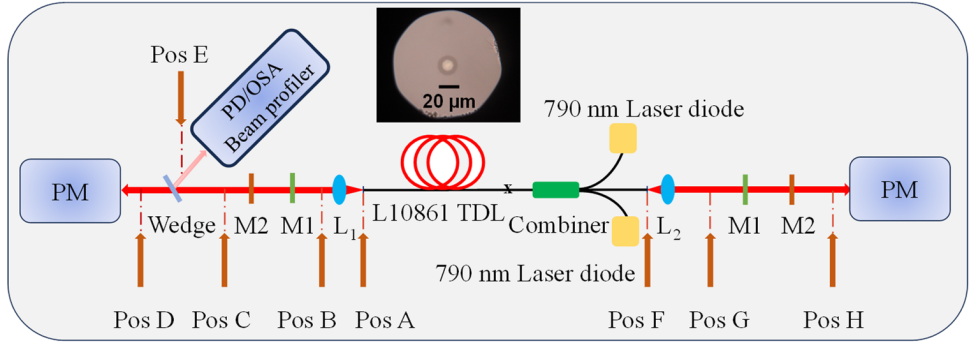

*Figure 1. Experimental setup of the all-fiber thulium-doped silica-based fiber laser system with two perpendicularly cleaved ends. TDF: thulium-doped fiber; PM: power meter; PD: photodetector; OSA: optical spectrum analyzer; $L_1$: AR-coated collimating lens; M1: short-pass filter, cut-off wavelength: 1600 nm; M2: long-pass filter, cut-on wavelength: 800 nm. A, B, C, D, E, F, G, H: Reference positions for measurements.*

The TDF cross-section is also shown in Fig. 1. The inner cladding is noticeably rounded from a proper octagon. A circular symmetry is known to reduce the pump absorption, which then improves with optimized coiling layouts [4]. However, the absorption in our TDF did not depend on the coiling layout, thus indicating that the shaping of the inner cladding was adequate. The TDF was laid down in a loose coil on the optical table with no special heat dissipation measures required. This demonstrates the effectiveness of QCW-pumping in simplifying the thermal management. Fig. 2 shows a TDF cladding transmission spectrum of white light, together with a spectrum of the 0.79-µm diode lasers, QCW at full power. The absorption peaks at 787.2 nm which matches well with the central emission wavelength of the diode lasers at full power. The operating pump absorption was ~8.6 dB. The core propagation loss at 1320 nm was measured with the cutback method to 0.15 dB/m. For this, the TDF was spliced between two pieces of SMF-28, and the transmission measured for two different lengths of the TDF.

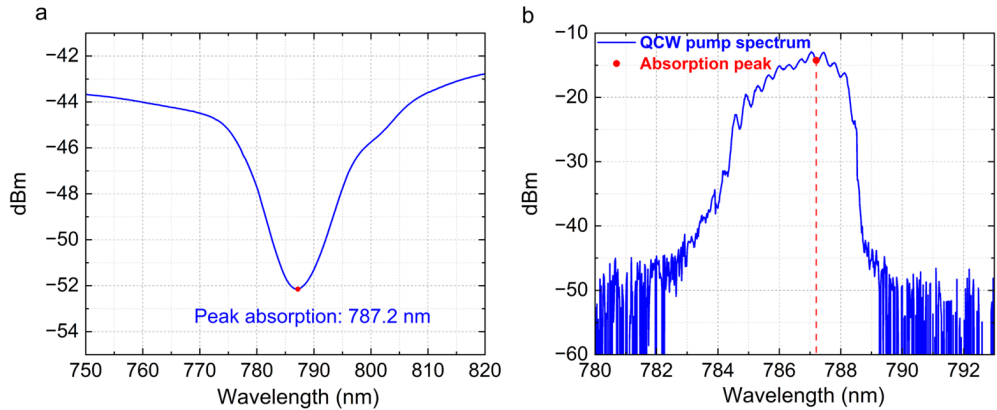

*Figure 2. (a) White-light cladding transmission spectrum of TDF. (b) QCW spectrum of diode lasers at full power.*

Two perpendicular fiber facets formed the laser cavity. The Fresnel reflectivities become 3.5% and 3.2% at 0.8 μm and 1.9 μm, respectively. At both ends of the cavity, AR-coated lenses ($L_1$ and $L_2$, AR-coating range 0.8~2 μm) collimated the output beams. Short-pass filters M1 (1600SP, Edmund Optics) and long-pass filters M2 (FELH0800, Thorlabs) separated the residual pump and generated (laser) beams. In addition, an uncoated wedge sampled the output in the far end (opposite the pump launch end) for low-power characterization of the output (e.g., temporal trace, spectrum, and beam quality). We compensate for the loss introduced by the wedge and filters in quoted output powers, which thus correspond to Position B and G in Fig. 1.

Diagnostic equipment of note included a Thorlabs DET10A/M biased photodetector (PD) with 350-MHz bandwidth, a Keysight InfiniiVision MSOX3054G oscilloscope with 500-MHz bandwidth, an Ophir 10A-SH and a Gentec XLP12-3S-H2-D0 thermal power meter, and a Thorlabs dual scanning-slit beam profiler (BP209VIS/M) with $M^2$ measurement system. For optical spectra, we used Ando AQ6317B and Yokogawa AQ6376 optical spectrum analyzers.

## 3. Laser results

Fig. 3 shows the instantaneous output power combined from both ends of the TDFL *vs.* instantaneous absorbed pump power. The total laser power exiting the fiber ends reached 198 W, combined at a longer wavelength of 1.91 μm and a shorter wavelength of 0.82 μm. The total slope efficiency with respect to absorbed pump power varied from 67% up to 200 W of absorbed pump power and 52% at the full 352 W of absorbed pump power. The slope efficiency with respect to launched pump power becomes 57% up to 240 W of launched pump power and 44% at full power. The short-wavelength threshold was 60 W. We did not investigate the characteristics below an instantaneous pump power of around 50 W because the LDD was unable to stably generate currents below 3.7 A, corresponding to ~50 W. The long-wavelength threshold was significantly lower than that.

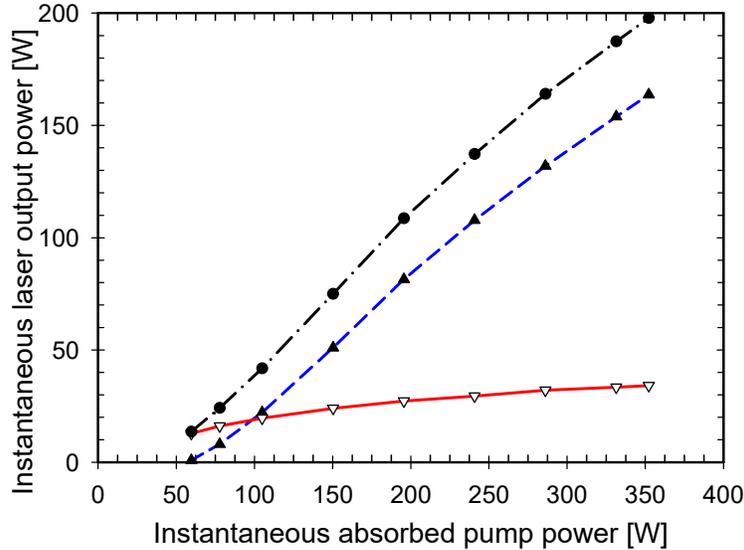

*Figure 3. Instantaneous output power combined from both ends of the TDFL vs. instantaneous absorbed pump power. Blue curve, dashed: short wavelength; red curve, solid: long wavelength; black curve, chain-dashed: total output power.*

Fig. 4 shows temporal traces of the pump and the short-wavelength TDFL output recorded with the biased PD at full power. The inset shows that the pump switches on within 10 μs (10%-90%). Short-wavelength relaxation oscillations occur approximately 5 μs after the pump power starts to rise, quickly dying out by the time the pump reaches full power (coincidentally or not). A slower pump-on time may suppress the relaxation oscillations, but this was not investigated.

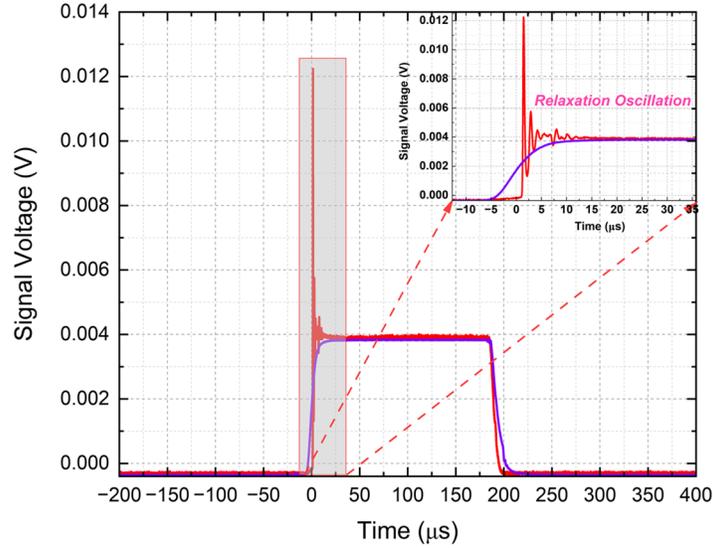

*Figure 4. Time-trace of short-wavelength TDFL output (red) and pump (purple) at full power. Inset: enlarged view of TDFL relaxation oscillations.*

The beam quality factor ($M^2$) of the short wavelength was measured in the far end of the TDFL with a scanning-slit beam profiler (Thorlabs) mounted on an $M^2$ measurement system with a motorized translation stage. The results are shown in Fig. 5, which were measured at Position E in Fig. 1. At the maximum pump power, the $M^2$-factor was determined to be 2.97 and 3.01 in orthogonal directions, based on a hyperbolic fit to the measured signal beam width at the $1/e^2$ intensity level.

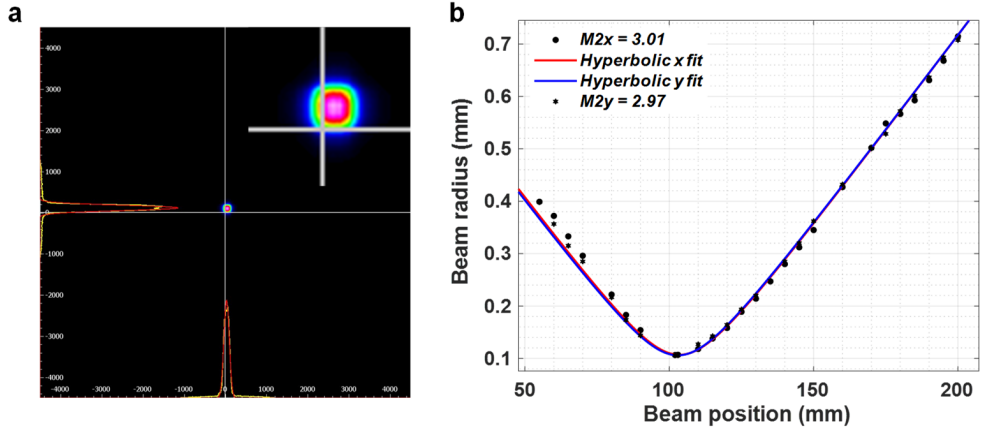

*Figure 5. Beam profiles measured in far end of the TDFL at the short wavelength. (a) Scanning-slit traces in orthogonal directions at focus. Inset: Synthetic 2D beam profile as reconstructed from the scanning-slit traces using an averaging technique. (b) Beam radii in orthogonal directions at different longitudinal positions at maximum power, together with hyperbolic fits.*

## 4. Conclusion

In conclusion, we have demonstrated a quasi-continuous-wave thulium-doped fiber laser in an all-fiber configuration. This produced up to 198 W of instantaneous output power when cladding-pumped by 408 W of power at 0.79 μm from two diode lasers. The slope efficiency with respect to absorbed pump power was 67% up to 200 W of absorbed pump power and 52% at the full 352 W of absorbed pump power. The diode lasers and thus the TDFL generated 0.2 ms pulses at 50 Hz for a duty cycle of 1%. The low duty-cycle allowed for simple water-free thermal management with the TDF lying in a loose coil on the optical table. Future work targets single-ended output, higher instantaneous output.

**Funding.** Air Force Office of Scientific Research (FA9550-17-1-0007), EPSRC EP/T012595/1 (LITECS)

**Disclosures.** The author declares no conflicts of interest.

**Data availability.** All data supporting this study are openly available from the University of Southampton repository at https://doi.org/10.5258/SOTON/D3308). [5].